\documentclass[12pt,a4paper]{article}

\usepackage{amsmath}
\usepackage{amsfonts}
\usepackage{amssymb}
\usepackage{graphicx}
\usepackage{color}
\usepackage{booktabs}
\usepackage{inputenc}
\usepackage[T1]{fontenc}
\usepackage{mathrsfs}
\usepackage{enumerate}
\usepackage{xcolor,url}
\usepackage{cancel,dsfont} 

\usepackage{amsmath}

\usepackage{slashed}

\def\hinvMpc{h\,{\rm Mpc}^{-1}}

\newcommand{\dd}{\partial}
\def\nn{\nonumber}

\def\PA{\slashed{P}}

\newcommand{\mpl}{M_{\rm Pl}}

\newcommand{\kmax}{k_{\rm max }}

\newcommand{\code}[1]{\texttt{#1}}

\usepackage[colorlinks,bookmarks]{hyperref}
\definecolor{linkblue}{rgb}{0,0,0.8}
\definecolor{linkgreen}{rgb}{0,0.5,0}

\hypersetup{pdfpagemode=UseNone, pdfstartview=FitH, linkcolor=linkblue,
 citecolor=linkgreen, urlcolor=linkblue}

\def\be{\begin{equation}}
\def\ee{\end{equation}}
\def\ba{\begin{eqnarray}}
\def\ea{\end{eqnarray}}
\def\knl{{k_{\rm NL}}}
\def\km{{k_{\rm M}}}
\def\knlr{{k_{\rm NL,\,R}}}

\font\BF=cmmib10
\def\k{{\hbox{\BF k}}}
\def\q{{\hbox{\BF q}}}

\begin{document}

\begin{center}

{\Large \bf {Taming redshift-space distortion effects in the EFTofLSS\\[0.3cm] and its application to data}
}
\\[0.7cm]

{\large Guido D'Amico${}^{1,2}$, Leonardo Senatore${}^{3}$, Pierre Zhang${}^{4,5,6}$, Takahiro Nishimichi${}^{7,8}$
\\[0.7cm]}
\end{center}

\begin{center}

\vspace{.0cm}

{\normalsize { \sl $^{1}$ Department of Mathematical, Physical and Computer Sciences,\\ University of Parma, 43124 Parma, Italy}}\\
\vspace{.3cm}

{\normalsize { \sl $^{2}$ INFN Gruppo Collegato di Parma, 43124 Parma, Italy}}\\
\vspace{.3cm}

{\normalsize { \sl $^{3}$ Institut fur Theoretische Physik, ETH Zurich,
8093 Zurich, Switzerland}}\\
\vspace{.3cm}

{\normalsize { \sl $^{4}$ Department of Astronomy, School of Physical Sciences, \\
University of Science and Technology of China, Hefei, Anhui 230026, China}}\\
\vspace{.3cm}

{\normalsize { \sl $^{5}$ CAS Key Laboratory for Research in Galaxies and Cosmology, \\
University of Science and Technology of China, Hefei, Anhui 230026, China}}\\
\vspace{.3cm}

{\normalsize { \sl $^{6}$ School of Astronomy and Space Science, \\
University of Science and Technology of China, Hefei, Anhui 230026, China}}\\
\vspace{.3cm}

{\normalsize { \sl $^{7}$ Center for Gravitational Physics, Yukawa Institute for Theoretical Physics, \\
Kyoto University, Kyoto 606-8502, Japan }}\\
\vspace{.3cm}

{\normalsize { \sl $^{8}$ Kavli Institute for the Physics and Mathematics of the Universe (WPI), UTIAS, \\
The University of Tokyo, Kashiwa, Chiba 277-8583, Japan }}\\
\vspace{.3cm}


\end{center}

\hrule \vspace{0.3cm}
{\small \noindent 
Former analyses of the BOSS data using the Effective Field Theory of Large-Scale Structure {(EFTofLSS)} have measured that the largest counterterms are the redshift-space distortion ones. This allows us to adjust the power-counting rules of the theory, and to explicitly identify that the leading next-order terms have a specific dependence on the {cosine of the} angle between the line-of-sight and the wavenumber of the observable, $\mu$. Such a specific $\mu$-dependence allows us to construct a linear combination of the data multipoles, $\PA$, where these contributions are effectively projected out, so that EFTofLSS predictions for $\PA$ have a much smaller theoretical error and so a much higher $k$-reach. The remaining data are organized in wedges in $\mu$ space, have a $\mu$-dependent $k$-reach because they are not equally affected by the leading next-order contributions, and therefore can have a higher $k$-reach than the multipoles. Furthermore, by explicitly including the highest next-order terms, we define a `one-loop+'  procedure, where the wedges have even higher $k$-reach. We study the effectiveness of these two procedures on several sets of simulations and on the BOSS data. The resulting analysis has identical computational cost as the multipole-based one, but leads to an improvement on the determination of some of the cosmological parameters that ranges from $10\%$ to $100\%$, depending on the survey properties.

\vspace{0.3cm}} 
\hrule

%
\newpage

\tableofcontents

\section{Introduction and Summary\label{sec:intro}}

\paragraph{Introduction} 
The analysis of the Full Shape (FS) of the BOSS galaxy power spectrum with the one-loop prediction from the Effective Field Theory of Large-scale Structure (EFTofLSS)  has obtained a measurement of all parameters in $\Lambda$CDM with just a Big Bang Nucleosynthesis~(BBN) prior~\cite{DAmico:2019fhj,Ivanov:2019pdj,Colas:2019ret} (see also~\cite{Philcox:2020xbv} for other prior choices and~\cite{DAmico:2019fhj} for a joint analysis with the BOSS bispectrum using the tree-level prediction). 
The FS analysis has been combined with BOSS reconstructed measurements and baryon acoustic oscillations (BAO) from eBOSS, as well as with supernovae redshift-distance or cosmic microwave background~(CMB) measurements, and this has further allowed us to put limits on the effective number of relativistic species, to bound the total neutrino mass and curvature, to constraint clustering and smooth dark energy~\cite{DAmico:2019fhj,Colas:2019ret,Ivanov:2019hqk,Philcox:2020vvt,DAmico:2020kxu,Chudaykin:2020ghx,DAmico:2020tty}. 
In particular, the FS analysis can help constrain models designed to ameliorate the Hubble tension as it provides measurements independent on the CMB or local distance ladders~\cite{DAmico:2020ods,Ivanov:2020ril,Niedermann:2020qbw,Smith:2020rxx,DAmico:2020tty}. Latest among these,~\cite{DAmico:2020tty} shows that the only known consistent model that can predict $w<-1$, which is clustering quintessence~\cite{Creminelli:2006xe}, does not ameliorate the Hubble tension and indeed $w$ is strongly constrained to be close to -1 even in this model~\footnote{We emphasize that even if clustering quintessence is a consistent quantum field theory, its discovery would represent a revolution of our understanding of quantum gravity, as in this context it is highly unexpected that one can have a consistent cosmological solution with $w<-1$.}.

We believe it is fair to say that these results were made possible by the development of the EFTofLSS, which is revealing itself to be a powerful instrument to extract cosmological information from Large-Scale Structure observations. 
A several-year long effort was necessary to bring this theory to the level where it can be applied to the data, and these efforts were conducted notwithstanding widespread skepticism on the actual usefulness of the EFTofLSS.  We therefore find it justified to add the following footnote where we acknowledge some of these developments, though not all intermediate results are used in the present analysis~\footnote{The initial formulation of the EFTofLSS was performed in Eulerian space in~\cite{Baumann:2010tm,Carrasco:2012cv}, and subsequently extended to Lagrangian space in~\cite{Porto:2013qua}. 
The dark matter power spectrum has been computed at one-, two- and three-loop orders in~\cite{Carrasco:2012cv, Carrasco:2013sva, Carrasco:2013mua, Carroll:2013oxa, Senatore:2014via, Baldauf:2015zga, Foreman:2015lca, Baldauf:2015aha, Cataneo:2016suz, Lewandowski:2017kes,Konstandin:2019bay}.
Accompanying theoretical developments were a careful understanding of renormalization~\cite{Carrasco:2012cv,Pajer:2013jj,Abolhasani:2015mra} (including rather-subtle aspects such as lattice-running~\cite{Carrasco:2012cv} and a better understanding of the velocity field~\cite{Carrasco:2013sva,Mercolli:2013bsa}), several ways for extracting the value of the counterterms from simulations~\cite{Carrasco:2012cv,McQuinn:2015tva}, and the non-locality in time~\cite{Carrasco:2013sva, Carroll:2013oxa,Senatore:2014eva}.
These theoretical explorations also include enlightening studies in 1+1 dimensions~\cite{McQuinn:2015tva,Pajer:2017ulp}.
In order to reproduce the Baryon Acoustic Oscillation (BAO) peak, an IR-resummation of the long displacement fields had to be performed, originating the so-called IR-Resummed EFTofLSS~\cite{Senatore:2014vja,Baldauf:2015xfa,Senatore:2017pbn,Lewandowski:2018ywf,Blas:2016sfa}.
An account of baryonic effects was presented in~\cite{Lewandowski:2014rca,Braganca:2020nhv}. The dark-matter bispectrum has been computed at one-loop in~\cite{Angulo:2014tfa, Baldauf:2014qfa}, the one-loop trispectrum in~\cite{Bertolini:2016bmt}, and the displacement field in~\cite{Baldauf:2015tla}.
The lensing power spectrum has been computed at two loops in~\cite{Foreman:2015uva}.
Biased tracers, such as halos and galaxies, have been studied in the context of the EFTofLSS in~\cite{ Senatore:2014eva, Mirbabayi:2014zca, Angulo:2015eqa, Fujita:2016dne, Perko:2016puo, Nadler:2017qto} (see also~\cite{McDonald:2009dh}), the halo and matter power spectra and bispectra (including all cross correlations) in~\cite{Senatore:2014eva, Angulo:2015eqa}. Redshift space distortions have been developed in~\cite{Senatore:2014vja, Lewandowski:2015ziq,Perko:2016puo}.
Neutrinos have been included in the EFTofLSS in~\cite{Senatore:2017hyk,deBelsunce:2018xtd}, clustering dark energy in~\cite{Lewandowski:2016yce,Lewandowski:2017kes,Cusin:2017wjg,Bose:2018orj}, and primordial non-Gaussianities in~\cite{Angulo:2015eqa, Assassi:2015jqa, Assassi:2015fma, Bertolini:2015fya, Lewandowski:2015ziq, Bertolini:2016hxg}. 
The exact-time dependence in the loop has been clarified in~\cite{Donath:2020abv,Fujita:2020xtd}. 
Faster evaluation schemes for the calculation of some of the loop integrals have been developed in~\cite{Simonovic:2017mhp}.
Comparison with high-fidelity $N$-body simulations to show that the EFTofLSS can accurately recover the cosmological parameters have been performed in~\cite{DAmico:2019fhj,Colas:2019ret,Nishimichi:2020tvu}.}. 

\paragraph{Summary} 

In the analysis of~\cite{DAmico:2019fhj,Ivanov:2019pdj,Colas:2019ret}, and confirmed by all the subsequent analysis, it was found that the size of the counterterms originating from redshift space distortions was particularly large, limiting the amount of data that could actually be used for a cosmological analysis by lowering the so-called $k$-reach of the theory, which is the maximum wavenumber up to which data can be {reliably} analyzed. In this paper, we first argue that the behavior of the redshift space counterterms can be understood by introducing a new momentum scale that controls the size of these terms, that we call $\knlr$, and which is lower than the one of the dark matter non-linearities, that is called $\knl$. This is the way the so-called Fingers of God make their appearance in the EFTofLSS. This understanding allows us to recognize that the most-enhanced terms have a specific $\mu$-dependence, with $\mu$ being the {cosine of the} angle between the observed wavenumber and the line of sight. Since in the EFTofLSS the $k$-reach is limited by the largest terms that appear at the perturbative order beyond the ones one uses, the specific $\mu$-dependence of the largest next-order terms allows us to define a linear combination of the data where these contributions are projected out, or at least up to a negligible amount. We call this linear combination $\PA\equiv P^{(D,\slashed{\mu^4},\slashed{\mu^6})}$, because it projects out the terms whose $\mu$-dependence is $\mu^4$ and $\mu^6$, and, up to a negligible amount, those whose dependence is $\mu^8$ and $\mu^{10}$. These are the functional forms associated to the largest next-order counterterms. The same combination accidentally almost projects out the $\mu^2$ components, further reducing the theoretical error of $\PA$. Since we work with the combination of three multipoles, we organize the two remaining ones in wedges in $\mu$ space, that we call $w_{1,2}$, and determine the $k$-reach of each of those by using the $\mu$-dependence of the leading next-order terms, so that the wedge {integrated over  lower values of $\mu$ has an higher $k$-reach than the one {integrated over} higher values. We dub this procedure as `one-loop', and apply it to both simulations and {BOSS} data. 

We also identify another procedure, that we dub `one-loop+'. Since we argue that the leading next-order terms from redshift space have a very simple functional form, we add them to the one-loop prediction of the EFTofLSS that is fitted to the data. In this way, the theoretical error associated to the two $\mu$-wedges (that are analyzed on top of $\PA$) is even smaller, so that it is possible to {reliably} analyze even more data. Similarly to the `one-loop' procedure, we apply `one-loop+' to both simulations and {BOSS} data.

\paragraph{Main Results:} 

We perform our analyses on four sets of data: on a large-volume simulation that we call `PT challenge', on two quite-large-volume sets of simulations populated with five HOD models, and that we call `Lettered Challenge', on a rescaled version of the `PT Challenge' simulation that we call `DESI-like', and finally on the BOSS data. The simulations are used to either estimate the theoretical error with great precision, or to estimate the $k$-reach of the EFTofLSS with the new methods in surveys such as DESI.  

The increase in the $k$-reach with respect to the multipoles depends on the cosmic variance, on the shot noise, and on the particular linear combination of the data we consider. Let us quote the results for the `one-loop+' procedure. For $\PA$, {we increase $k_{\rm max}$ on the PT challenge by a factor of 2.5, by 2.3 for DESI-like and by 1.5 for BOSS}. The $k$-reach of the wedge with the smaller values of $\mu$, $w_1$, is increased by a factor of 1.8 for PT challenge, 1.7 for DESI-like, and {negligibly modified} for BOSS. The $k$-reach of $w_2$ is instead decreased with respect to the multipoles, in order to keep the value of the theoretical error constant in all {three data combinations we use}. The `one-loop' procedure has the same $k$-reach for the bin $\PA$ as the `one-loop+',  and a slightly lower, but still significant, {increase} in the $k$-reach of $w_1$.

Similarly, the increase in the precision in the determination of the cosmological parameters depends on the cosmic variance and the shot noise of the data. Quoting results for `one-loop+', for PT challenge we increase the precision by $60\%$ or $70\%$ for parameters such as $\Omega_m$, $A_s$ and $h$, and $105\%$ for $n_s$. The improvements for `one-loop' procedure are smaller by about $10\%$ or $15\%$, depending on the cosmological parameters. Unfortunately, the improvements are not as remarkable for a DESI-like survey, where we obtain an improvement that is about $25\%$ for $\Omega_m$ and $h$, 10\% for $n_s$ and $5\%$ on $A_s$, with similar results for the `one-loop' procedure. Finally, on the BOSS data, we find that by using the `one-loop+' procedure, we improve the determination of $\Omega_m$ by $10\%$, while we have marginal improvements for the other cosmological parameters. Instead, 'one-loop' does not lead to any improvement on BOSS data.

To summarize, we designed an analysis to mitigate the size of the theory error associated to redshift-space distortions in the EFTofLSS.  
Compared to the standard fit to multipoles, this new analysis leads to a significantly more accurate and precise determination of the cosmological parameters. 
This is made possible as we analyze a much larger number of modes, made accessible thanks to two improvements. 
First, the multipoles are rotated into new linear combinations, $\PA$ and $w_{1,2}$. 
$\PA$ is designed such that the associated theory error is strongly mitigated, allowing for a much higher $k$-reach,
while the $k_{\rm max}$'s in $w_{1,2}$ are pushed as far as possible with a controlled theory error depending on their specific $\mu$-range  (`one-loop' procedure). 
Second, the leading  next-order EFT-counterterms have been identified and can be included in order to further increase the $k$-reach in the $\mu$ direction (`one-loop+' procedure). 
} 
Given that performing this kind of analysis has practically identical computational cost as the standard multipole-based analysis, we conclude that we believe there is no reason why the {`one-loop'} or the `one-loop+' procedures should not be routinely used in future analyses.

We end this summary of the main results with a note of warning. It should be emphasized that in performing the analysis presented here for the BOSS data, as well as the preceding ones using the EFTofLSS by our group~\cite{DAmico:2019fhj,Colas:2019ret,DAmico:2020kxu,DAmico:2020tty,DAmico:2020ods}, we have assumed that the observational data are not affected by any unknown systematic error, such as, for example, line of sight selection effects or undetected foregrounds. In other words, we have simply analyzed the publicly available data for what they were declared to be: the power spectrum of the galaxy density in redshift space. Given the additional cosmological information that the theoretical modeling of the EFTofLSS allows us to exploit in BOSS data, it might be worthwhile to investigate if potential undetected systematic errors might affect our results. We leave an investigation of these issues to future work.

\paragraph{Public Code} The redshift-space one-loop galaxy power spectra in the EFTofLSS are evaluated using PyBird: Python code for Biased tracers in ReDshift space~\cite{DAmico:2020kxu}~\footnote{\href{https://github.com/pierrexyz/pybird}{https://github.com/pierrexyz/pybird}}.
The linear power spectra are evaluated with the CLASS Boltzmann code~\cite{Blas_2011}~\footnote{\href{http://class-code.net}{http://class-code.net}}.
The posteriors are sampled using the MontePython cosmological parameter inference code~\cite{Brinckmann:2018cvx, Audren:2012wb}~\footnote{\href{https://github.com/brinckmann/montepython\_public}{https://github.com/brinckmann/montepython\_public}}.
The triangle plots are obtained using the GetDist package~\cite{Lewis:2019xzd}.

\section{Theoretical Considerations}

\subsection{Estimates of the scales governing the EFTofLSS expansion}

In the EFTofLSS, the various EFT-parameters that are present in the theory account for the effect of short-distance physics at long distances. In practice, in trying to solve perturbatively the equations that define the various operators over which we take expectation values, we encounter several terms whose evaluation is UV-sensitive, {\it i.e.} their evaluation requires knowledge of short-distance physics. Luckily, we can parametrize our ignorance of short-distance fluctuations by expanding in a series of counterterms that can be evaluated perturbatively. This perturbative expansion can however have various expansion parameters, and some UV contributions can be larger than others. This is what seems to happen in the EFTofLSS, as we are now going to explain.

In the original data analysis that provided a measurement of cosmological parameters from Large-Scale Structure data~\cite{DAmico:2019fhj,Colas:2019ret}, as well as in {the comparison of} the EFTofLSS predictions with simulations, several of these parameters have been measured. At one-loop order, schematically, the prediction of the power spectrum in redshift space contains terms such as
\be
P_{g,r}(k,\mu,t)\supset c_{ct}(t) \frac{k^2}{\knl^2} P_{11}(k,t)+c_{r}(t) \mu^2\frac{k^2}{\knl^2} P_{11}(k,t)\ ,
\ee
where $P_{11}(k)$ is the matter linear power spectrum, {$\mu = \hat{k} \cdot \hat{z}$} is the {cosine of the} angle of the wavenumber $\vec k$ with the line of sight, $\hat z$, and $\knl(t)$ is the time-dependent scale of dark-matter non-linearities. 

In~\cite{DAmico:2019fhj,Colas:2019ret}, with $\knl=0.7\hinvMpc$, it was measured that $c_{ct}\sim 1$,  while $c_r\sim 8$ (see Table 3 of~\cite{DAmico:2019fhj}). Now, the origin of the EFT-parameters that do not depend on $\mu$, such as $c_{ct}$, is quite different than {the one of} those that depend on $\mu$, $c_r$. In fact, the terms that depend on $\mu$ are counterterms of UV-sensitive operators that appear when we {transform} our predictions from configuration space to redshift space~\cite{Senatore:2014vja}. The particular terms under question are expectation values of product of velocities at the same location. For example, the term in $c_r$ arises when we write~\cite{Senatore:2014vja} 
\be\label{eq:velocity1}
\langle v_i(\vec x,t) v_j(\vec x,t)\rangle=\left(\frac{aH}{\knl}\right)^2 \left[ c_1(t) \delta_{ij} + \left( c_{2}(t) \delta_{ij} + c_{3}(t) \frac{\dd_i\dd_j}{\nabla^2} \right) \delta(\vec x,t) + \tilde c_r(t) \frac{\dd_i\dd_j}{\knl^2}\delta(\vec x,t)+\ldots \right] \, ,
\ee
with $c_r = c_1 + c_2 + \dots$, $\delta$ is the dark matter density and $\ldots$ here and everywhere else in this paper represents a list of all possible terms allowed by the symmetries, written in an expansion in the size of the fluctuations and the size of the derivatives, $\dd/\knl$ (for the purpose of this discussion, we here neglect the fact that the counterterms are non-local in time~\cite{Senatore:2014eva}, a fact that is however taken into account in the formulas we use to compare with data).
On the other hand, the term in $c_{ct}$ comes from either expanding the expectation value of the matter effective stress tensor or from the derivative expansion of the galaxy fields in terms of the dark matter fields:
\ba\label{eq:stess_tensor1}
&&\frac{1}{\rho_0(t)}\langle\tau_{ij}(\vec x,t)\rangle=\left(\frac{a H}{\knl^2}\right)\left(c_0 \delta_{ij}+c_{s}^2(t)  \delta(\vec x,t)+c_{4}(t) \frac{\dd^2}{\knl^2} \delta(\vec x,t)+\ldots\ \right),\\ 
\label{eq:galaxy_derivative1}
&&\delta_g(\vec x,t)=b_1(t) \delta^{(1)}(\vec x,t)+ c_{ct,2}(t) \frac{\dd^2}{\km^2}\delta(\vec x,t)+\dots \ ,
\ea
where $\rho_0\sim H^2\mpl^2$ is the background energy density, $\km$ is the {typical wavenumber associated to galaxy size} and $\delta_g$ is the galaxy overdensity. From here we get $c_{ct}\equiv c_s^2+c_{ct,2}$. 

In the analysis of~\cite{DAmico:2019fhj,Colas:2019ret}, it was taken that $\knl=\km=0.7 \hinvMpc$. Now, the fact that on observations we measure that $c_{r}\gg c_{ct}\sim 1$ suggests that $\langle v_i(\vec x,t) v_j(\vec x,t)\rangle$ is large and defines a new {length} scale which is longer than the non-linear scale. We can therefore improve (\ref{eq:velocity1}) by introducing a new scale {$\knlr\sim \knl/\sqrt{8}$}, and writing
\be\label{eq:velocity2}
\langle v_i(\vec x,t) v_j(\vec x,t)\rangle=\left(\frac{aH}{\knlr}\right)^2\left[c_1(t) \delta_{ij} +  \left( c_{2}(t) \delta_{ij} + c_{3}(t) \frac{\dd_i\dd_j}{\nabla^2} \right) \delta(\vec x,t) + \tilde c_r(t) \frac{\dd_i\dd_j}{\knl^2}\delta(\vec x,t)+\ldots\right] \, .
\ee
In this way, the new $c_r$ is measured to be order one. The fact that $\langle v_i(\vec x,t) v_j(\vec x,t)\rangle$ is large is the manifestation, in the EFTofLSS, of what the Large-Scale Structure community often refers to as the {large size} of the Finger-of-God effects. 

Now that we have introduced a new scale, $\knlr$, we need to address where else it enters. Where should we replace the factors of $\knl$ with $\knlr$ in the above equations? Nicely, we can answer this questions without relying on any astrophysical insight, but using the measurements and the assumption that unitless numerical coefficients should be {of} order one. 

Let us start with the stress tensor: should we replace $\knl$ in the first factor in (\ref{eq:stess_tensor1})? The answer is `no'. In fact, if this were the case, the dark matter velocity counterterm would take the form $c_s^2\dd^2\delta/\knlr^2$. But $c_s^2$ has been measured to be {of} order one in many dark matter simulations after normalizing with $\knl$. For example, eq.~(20) of~\cite{Foreman:2015lca} gives $c_s^2\simeq 0.3$, which is order one, while replacing $\knl$ with $\knlr$ would give $c_s^2\to 0.03$, which can hardly be thought of as an order one number. Let us make an additional observation for the stress tensor. If we look at the expression of the stress tensor in terms of UV terms, it reads, schematically~\cite{Baumann:2010tm}, $\tau_{ij}(\vec x)\sim \rho(\vec x) \left(v_i(\vec x)v_{j}(\vec x)+\delta_{ij} \Phi(\vec x)\right)$, with $\Phi$ being the gravitational potential. The presence of $v_iv_j$ might raise the suspicion that perhaps $\langle\tau_{ij}\rangle$ should be similarly enhanced by $1/\knlr^2$. However, this is a different operator, and so can have a different value; and in fact, Ref.~\cite{Baumann:2010tm} first pointed out that there is a very strong cancellation between the kinetic and the potential energy in the stress tensor, so that the effective pressure is small. Measurements are so far indicating that a cancellation occurs even at the level of the $c_s^2$ term (which, physically, represents the response of the stress tensor to a deformation induced by a $\delta$ fluctuation).

Let us move to the derivative expansion of the galaxies in~(\ref{eq:galaxy_derivative1}). Since we measured $c_s^2\simeq 1$ in matter simulations, we can say that the observational data of~\cite{DAmico:2019fhj,Colas:2019ret} measured $c_{ct,2}$ to be {of} order one as well. If instead we replaced $\km$ with $\knlr$ {, $c_{ct,2}$ would be $\sim 0.1$.} Therefore, though $\km$ is a different scale than $\knl$, we find no evidence that it cannot be assumed to be similar to $\knl$. We conclude that the derivative expansion of galaxies ({at least} of the kind observed by BOSS) is controlled by $\knl$, which is in agreement with what we would expect on astrophysical grounds.

Next, we move to the derivative expansion of the stress tensor. Should the scale suppressing the derivatives in~(\ref{eq:stess_tensor1}) be replaced with $\knlr$? Let us use measurements again. The higher-derivative term for dark matter was measured in~\cite{Foreman:2015lca} (eq.~(20) again), obtaining the order one number $c_4\simeq 2$ (after adjusting $\knl\to 0.6\hinvMpc$ as the measurements of~\cite{Foreman:2015lca} were done at redshift zero). On the contrary, if we were to use $\knlr$ for the derivative expansion of the stress tensor, we would obtain $c_4\to16$, which is not an order one number. We conclude that we should use $\knl$ in the derivative expansion for the stress tensor. This is again in agreement with what we would have expected on astrophysical grounds.

A final comment about the stochastic terms.
The measurements in simulations and on the data of [1, 3] do not show any particular enhancement of those terms.

In summary, in agreement with expectation from astrophysical considerations, we find that measurements in simulations and observations seem to indicate that the derivative expansion of any operator is governed by $\knl$. Similarly, the size of $\tau_{ij}$ is of order $H^2/\knl^2$. On the other hand, expectation values of the velocity field are enhanced and governed by $v^{n}\sim \left(H/\knlr\right)^{n}$. This tells us that, in the EFTofLSS, the redshift space counterterms are, in a sense, the largest ones.

\subsection{{Taming} Redshift Distortion Effects}

We now proceed to study how we can use our acquired knowledge to mitigate the effects of redshift space distortions. In the EFTofLSS, when making predictions at a given order, we have an estimate of the next-order corrections, which are the ones that, growing at shorter wavenumbers, limit the  $k$-reach of the theory, {\it i.e.} the maximum wavenumber at which predictions can be trusted. In this paper, we will compare the one-loop predictions of the EFTofLSS for the power spectrum against data. The $k$-reach is dictated by two two-loop contributions, which we are now going to explore. 

At two-loop order, we can estimate the size of the various contributions by looking at the tree-level counterterms of the same order, and assuming coefficients of order one once the scaling rules described above are applied. At two-loop order, we expect counterterms of the form $k^4P_{11}$. The largest ones are expected to be the ones suppressed by $\knlr^4$. Interestingly, these terms come with a definite $\mu$-dependence. In fact, from the scaling {discussed in the previous} subsection, in order to get four factors of $1/\knlr$, we need the counterterms for four velocities, as anything else will just {bring} factors of $1/\knl$. This means that $1/\knlr^4$ can come from the counterterm to this particular `51' contribution at two-loop order (following the formalism of~\cite{Senatore:2014eva}):
\ba\label{eq:counter1}
&&P_{51}(k,\mu)\supset \frac{k_z}{aH} \frac{k_z}{aH}\frac{k_z}{aH}\frac{k_z}{aH}\langle\left[v^z v^z v^z v^z \left(\delta_{g}-\frac{4!}{5!}i\frac{k_z}{a H}v_z\right) \right]^{(5)}_{\vec k} [\delta_{g,r}]^{(1)}_{\vec k'}\rangle'\\ \nn
&&\qquad\to \left(\frac{k}{\knlr}\right)^4\mu^4 \left(b_1+\frac{1}{5}f\mu^2\right) \left(b_1+f\mu^2\right) P_{11}(k)\ ,
\ea
where $[{\cal{O}}]^{(n)}_{\vec k}$ means that we take the wavenumber $\vec k$ of operator ${\cal{O}}$, evaluated at order $n$ in perturbation theory, $\langle\ldots \rangle'$ means that we drop the momentum-conserving Dirac $\delta$-function, and $\delta_{g,r}$ is the galaxy overdensity in redshift space. By the symbol $\to$ we mean that the diagram has the counterterm on the right of the arrow. To obtain this result, we simply substituted $\langle v^i v^j\rangle\to\delta^{ij} \left(\frac{a H}{\knlr}\right)^2$. The factor of $4!/5!$ comes from the Taylor expansion of the exponential for the redshift space distortions effect (notice that we are neglecting the $1/4!$ in front of this term).

It is important to stress the following point. The largest {contributions} come from terms enhanced by $1/\knlr^4$. These terms are larger than the loops of the theory, which are are just suppressed by the dark-matter non-linear scale, $\knl$, that enters in the power spectrum. In particular, this applies to the UV limit of the loops, which is not enhanced by $1/\knlr^4$. It is just the UV counterterm that is enhanced.  This in particular implies that we know the $k$ and $\mu$ dependence of the higher-order counterterms that are enhanced.

Before going on, let us identify the remaining maximally-enhanced counterterm. It turns out there is only one.  We consider the $P_{51}(k,\mu)$ term which generates from this:
\ba\label{eq:counter2}
&&P_{51}(k,\mu)\supset \frac{k_z}{aH} \frac{k_z}{aH}\frac{k_z}{aH}\frac{k_z}{aH}\langle\left[v^z v^z v^z v^z \right]^{(5)}_{\vec k} [\delta_{g,r}]_{\vec k'}\rangle'\to\\ \nn
&&\qquad\to \frac{k_z}{aH} \frac{k_z}{aH}\frac{k_z}{aH}\frac{k_z}{aH}\left(\frac{a H}{\knlr}\right)^4 \langle\left[\dd_z \dd_z\Phi \right]_{\vec k} [\delta_g]_{\vec k'}\rangle' \to \left(\frac{k}{\knlr}\right)^4\mu^6 \left(b_1+f \mu^2\right) P_{11}(k)
\ea
where in the second step we substituted $\langle v^i v^j\rangle\to\delta^{ij} \left(\frac{a H}{\knlr}\right)^2$ and $\langle v^i v^j\rangle\to \left(\frac{a H}{\knlr}\right)^2\frac{\dd_i\dd_j \Phi}{(aH)^2}$.

It is easy to realize that the scaling assignments of the former subsection do not allow any other term enhanced by $1/\knlr^4$. We therefore obtain two maximally-enhanced terms: $\sim\mu^4 (k/\knlr)^4P_{11}$ and $\sim\mu^6 (k/\knlr)^4P_{11}$. There are other enhanced (but not maximally enhanced) counterterms that include a lower-{order} $\mu$-dependence~\footnote{To obtain a given power of $(k/\knlr)^{2n}$, with $n$ large, one needs a factor of at least order $\mu^n$, suggesting that the effective expansion parameter of the maximally enhanced terms is $\knlr/|\mu|$. The Finger of God is clearly pointing at us or away from us.}. For example, by evaluating at one-loop order the one-loop counterterm: 
\ba
&&P_{31,{\rm counter,\; one-loop}}(k,\mu)\supset \frac{k_z}{aH} \frac{k_z}{aH}\langle\left[v^z v^z \right]^{(3)}_{\vec k} [\delta_g]_{\vec k'}\rangle'\to\\ \nn
&&\qquad \to \frac{k_z}{aH} \frac{k_z}{aH}\langle\left[\delta^{zz}\frac{\dd^2}{\knl^2} \delta \right]_{\vec k} [\delta_g]_{\vec k'}\rangle'\to b_1\mu^2\left(\frac{k}{\left(\knl\knlr\right)^{1/2}}\right)^4P_{11}(k)\ .
\ea
So, we see that at the level of $\mu^2$ terms, we obtain terms enhanced by $1/(\knlr^2\knl^2)$. However, in this case the functional dependence in $k$ is not known, as the counterterm is not parametrically separated in size from the term that we obtain by evaluating the loop in $\langle\left[v^z v^z \right]^{(3)}_{\vec k} [\delta_g]_{\vec k'}\rangle'$, which would give rise to a more complex $k$-dependence.
Finally, at order $\mu^0$, we do not obtain any $\knlr$-enhanced terms.

{\bf Procedure `one-loop':} This {discussion suggests} the following two procedures for enhancing the $k$-reach of the theory. The first procedure, that we call `procedure `one-loop'', is the following. Assuming, without loss of generality, that the measurements of the data are provided in multipoles of the angle $\mu$, we can construct a linear combination of the multipoles that removes the largest two-loop contributions. As we said, these are the ones enhanced by $1/\knlr^4$. The resulting linear combination will have a smaller theoretical error and therefore {a} higher $k$-reach. Since these terms have a specific $\mu$-dependence, the linear combination can be constructed independently of the actual size of the terms that we wish to eliminate.  Given enough multipoles, one could also remove the $\mu^2$ contribution, reaching an even higher $k$-reach, as in this case all the {subleading contributions} enhanced by $1/\knlr^2$ would also be eliminated. Since the data we have at our disposal provide only three multipoles, we decide to remove only the $\mu^6$ and $\mu^4$. This means that we are not removing exactly a contribution with the $\mu$-dependence of (\ref{eq:counter1}), which has also a $\mu^8$ dependence, upon which we will comment shortly. Writing the data as
\be
P^{\rm (D)}(k,\mu)=\sum_{\ell=0,2,4} P_\ell^{\rm (D)}(k) {\cal{L}}_\ell(\mu)\ ,
\ee
with $ {\cal{L}}_\ell(\mu)$ being the Legendre polynomial of order $\ell$, the resulting linear combination that projects out  the $\mu^4$ and $\mu^6$ dependence, and that we call $P^{\rm (D,\slashed{4},\slashed{6})}$, or for brevity, $\PA$, is given by the following linear combination:
\be\label{eq:Pslash}
\PA(k)=P_{\ell=0}^{\rm (D)}(k)-\frac{3}{7}P_{\ell=2}^{\rm (D)}(k)+\frac{11}{56}P_{\ell=4}^{\rm (D)}(k)\ .
\ee

While this linear combination does not remove a term proportional to $\mu^8$, which is expected to be present proportional to $1/\knlr^4$, notice that a term of the form $1\cdot\mu^8$ gets projected in $\PA$ as a small number equal to $5/1287\simeq 0.004$, which makes this contribution quite negligible. Evidently, $\PA$ approximately projects away also a $\mu^8$ component. Similar considerations apply to {the prefactor of an eventual} $\mu^{10}$ term: upon projection, it gets suppressed by a factor of $1/143\simeq 0.007$.

The two-loop contribution given by  the $\PA$ combination of data scales therefore in size approximately as $(k/\knl)^4P_{11}$ and as $\mu^2(k/(\knl \knlr)^{1/2})^4 P_{11}$. Notice, again, that, unlike the contributions in $1/\knlr^4$, the $k$-dependence of these {contributions}  is not known, as there are comparable contributions from the loops. While naively the contribution to the theoretical error from the term in $1/\knlr^2$ is the dominant one, the combination in $\PA$ highly suppresses also the combination that goes as $\mu^2$. A term of the form $1\cdot\mu^2$ gets projected in $\PA$ as s small number equal to $1/21$. This is such a large suppression to make the leading theoretical error the one in $1/\knl^4$. Therefore, if $\slashed{\sigma}^{\rm (D)}(k)$ is the observational error of the combination $\PA$, one can impose the theoretical error to be a fraction equal to $\epsilon$ of the observation error (say, $\epsilon=1/3$):
\be
\epsilon=\frac{\left(\frac{k}{\knl}\right)^4 P_{11}(k)}{\slashed{\sigma}^{\rm (D)}(k)}\ ,
\ee 
and determine in this way the $k$-reach of $\PA$, $k_{\rm max}^\slashed{P}$, by solving the above equation. Notice that we have set the ${\cal{O}}(1)$ numerical coefficient in front of the theoretical error to be equal to one, which is inaccurate, but we cannot do better than this. In practice, we will determine the $k$-reach directly with simulations, though this formula enlightens how the $k$-reach is increased by decreasing the theoretical error. 
 
What to do of the remaining two linear independent combinations of the data? It does not appear worthwhile to eliminate either the $\mu^4$ or the $\mu^6$, as the theoretical error would not be parametrically improved (unless we eliminate both, but that would give $\PA$). Given a configuration of data in the $\mu$ direction, $P^{\rm (D)}(k,\mu)$, the associated theoretical error is, roughly, $\frac{k^4}{\knl^4}\left(1+\left(\mu^4+\mu^8\right)\left(\frac{\knl}{\knlr}\right)^4\right)P_{11}$, where for simplicity we dropped the $\mu^2 k^4/(\knl^2\knlr^2)P_{11}$ contribution, as it does not matter much. Again, lacking a better procedure, we have put the prefactors to the expressions of the theoretical errors to one. This allows us to write a nice formula for the $k$-reach as a function of $\mu$, $k_{\rm reach}(\mu)$, in terms of $k_{\rm max}^\slashed{P}$:
\be\label{eq:kreachbins}
k_{\rm reach}(\mu)=k_{\rm max}^\slashed{P}\cdot\left(\frac{1}{\left(1+\left(\mu^4+\mu^8\right)\left(\frac{\knl}{\knlr}\right)^4\right)}\right)^{1/4} \cdot\left(\frac{\sigma(k,\mu)}{\slashed{\sigma}^{\rm (D)}(k)}\right)^{1/4}\ .
\ee
Therefore, we divide the resulting two combinations of data in wedges in $\mu$ space, one in $\mu\in[0,1/2]$, that we call $w_1$, and the other $\mu\in(1/2,1]$, that we call $w_2$, and denote the associated $k$-reach with {$k_{\rm max}^{w_1}$ and $k_{\rm max}^{w_2}$}, determined in terms of $k_{\slashed{P}}$ using the formula (\ref{eq:kreachbins}).  It is now possible to give a more mathematical interpretation of the combination $\PA$. This is obtained by noticing that, given the two wedges, the linear combination $\PA$ is the one that maximizes the signal to noise in the limit in which $\knl/\knlr$ is very large~\footnote{We also notice that, if we were to divide the data in three wedges instead of two, and neglected the contribution of $\PA$, the theoretical error of each wedge would receive contributions also from the terms in $\mu^4(k/ \knlr)^4 P_{11}$, and $\mu^6(k/ \knlr)^4 P_{11}$. However, for the wedge with $\mu\in[0,1/3]$, the smallness of $\mu$ is such that the error is actually dominated by the real space one, $\left(k/\knl\right)^4P_{11}$, as it is for $\PA$. In fact, we noticed that the correlation of $\PA$ with the wedge with $\mu\in[0,1/2]$ is already very high. We therefore expect that one would obtain similar results using a suitable choice of three wedges.}.

In the literature (see~\cite{Scoccimarro:2004tg}) there is also another choice for a data combination which suppresses redshift-space distortion terms, namely $Q(k) = P_{\ell=0}(k) - \frac{1}{2} P_{\ell=2}(k) + \frac{3}{8} P_{\ell=4}(k)$. This is the real-space power spectrum for the case in which only multipoles up to $P_{\ell=4}$ are present.
With respect to our $\slashed{P}$, $Q(k)$ has no projection of the $\mu^2$ component, but it keeps the projection of the $\mu^6$ component which projects onto it with a coefficient $\frac{5}{231} \simeq 0.02$. The $\slashed{P}$ combination instead keeps the projection of the $\mu^2$ component, suppressed by $1/21 \simeq 0.05$. However, as explained above, we expect that, when looking at the NNLO terms $\propto k^4$, the $\mu^2$ terms will be suppressed with respect to the $\mu^4$, $\mu^6$ ones by a factor $k_{\rm NL}^2 / k_{\rm NL,R}^2 \simeq 8$, thus also this term is almost cancelled out in $\slashed{P}$.
We have checked that the correlation between $\slashed{P}$ and $Q(k)$ is very high, larger than $0.95$. We expect that one would obtain similar results using $Q(k)$ in place of $\slashed{P}$.~\footnote{Note added in print. \cite{Ivanov:2021fbu}, submitted on the arXiv in coordination with our paper, uses the $Q(k)$ variable. They find similar gains in error bars as we do, although a precise comparison cannot be made because of different analysis choices.}

So far, we have worked assuming that the available prediction for the EFTofLSS was at one-loop order: this is what determined the theoretical error. Generalizations to the case {for which} the EFTofLSS prediction is available at higher orders or for higher $n$-point functions {are} straightforward.  This completely defines what we call `procedure `one-loop''.

{\bf Procedure `one-loop+':} All of these considerations suggest that there is a rather straightforward way to improve the predictions of the EFTofLSS at one-loop order. We argued that the maximally-enhanced theoretical error has the simple form of $\sim \mu^4 (k/\knlr)^4P_{11}$ and $\sim\mu^6 (k/\knlr)^4P_{11}$ (given precisely in (\ref{eq:counter1}) and (\ref{eq:counter2})), as it comes from the counterterms and not from the loops. Since the functional form is completely known, one can add them to the prediction at one-loop order, to obtain a sort of 1-loop+ order:
\ba
&&P_{\rm EFTofLSS,\ 1-loop+}(k,\mu)=P_{\rm EFTofLSS,\ 1-loop}(k,\mu)+\\ \nn
&&\quad+c_{r,4}\,\mu^4 \left(b_1+\frac{1}{5}f\mu^2\right) \left(b_1+f\mu^2\right)  \left(\frac{k}{\knlr}\right)^4P_{11}(k)
+c_{r,6} \,\mu^6 \left(b_1+f\mu^2\right)  \left(\frac{k}{\knlr}\right)^4P_{11}(k)\ .
\ea
Notice that since we have two wedges, this {procedure amounts} to add two independent terms of the form $k^4 P_{11}$, with the proper prior, to each wedge.
Now, the theoretical error in the $\mu$ direction is dominated by the terms in $\mu^{2 n} \left(\frac{k}{(\knl\knlr)^{1/2}}\right)^4 P_{11}$. Therefore, while the $k$-reach of $\PA$ is unaltered, the one of the $\mu$-wedges is raised according to the following formula: 
\be\label{eq:kreachbins2}
k_{\rm reach+}(\mu)=k_{\rm max}^\slashed{P}\cdot\left(\frac{1}{\left(1+3\left(\mu^4+\mu^6+\mu^{12}\right)^{1/2}\left(\frac{\knl}{\knlr}\right)^2\right)}\right)^{1/4} \cdot\left(\frac{\sigma(k,\mu)}{\slashed{\sigma}^{\rm (D)}(k)}\right)^{1/4}\ ,
\ee
where we added the various $\mu$-dependent contributions in quadrature and the numerical factor `3' has been chosen by optimizing against simulations. This concludes the explanation of procedure `one-loop+', whose generalization to higher orders is straightforward. We now proceed to investigate how both procedures perform on the data.

\section{Results}

\subsection{Likelihood and Priors}\label{sec:lkl}
Having discussed the combination of multipoles which mitigate the effects of redshift-space distortions, we now turn to the data analysis.
We construct our new data combinations, that we will denote $\PA + w_{1,2}$ in what follows, and refer to it simply as `wedges', starting from the measurements of three multipoles (monopole, quadrupole, hexadecapole), that we will denote, as earlier, as $P_{\ell=\{0,2,4\}}$.   
The likelihood is a Gaussian, with the new combinations of data and theory model being a linear transformation of the multipoles (and the data covariance given by a bilinear transformation).
{Explicitly, the transformation is~\footnote{We take this matrix to define our $w_1$, $w_2$ combinations. They correspond to the wedges of $P(k, \mu)$ if we approximate $P(k, \mu) = P_{\ell=0}(k) \mathcal{L}_0(\mu) + P_{\ell=2}(k) \mathcal{L}_2(\mu) + P_{\ell=4}(k) \mathcal{L}_4(\mu)$, where $\mathcal{L}_{\ell}$ are Legendre polynomials.}:
\begin{equation}
    \begin{pmatrix}
        \PA \\
        w_1 \\
        w_2
    \end{pmatrix} = 
    \begin{pmatrix}
        1 & -\frac{3}{7} & \frac{11}{56} \\
        1 & -\frac{3}{8} & \frac{15}{128} \\
        1 & \frac{3}{8} & - \frac{15}{128}
    \end{pmatrix}
    \begin{pmatrix}
        P_{\ell=0} \\
        P_{\ell=2} \\
        P_{\ell=4}
    \end{pmatrix} \, .
\end{equation}
}

The theory prediction is given by the galaxy power spectrum in redshift space at one-loop order in the EFTofLSS as described in~\cite{Perko:2016puo,DAmico:2019fhj,Nishimichi:2020tvu}.
The evaluation is performed using \code{PyBird}~\cite{DAmico:2020kxu}.
Explicitly, the one-loop redshift-space galaxy power spectrum reads:
\begin{align}\label{eq:powerspectrum}
P_{g}(k,\mu) & =  Z_1(\mu)^2 P_{11}(k)  \\
& + 2 \int d^3q\; Z_2(\q,\k-\q,\mu)^2 P_{11}(|\k-\q|)P_{11}(q) + 6 Z_1(\mu) P_{11}(k) \int\, d^3 q\; Z_3(\q,-\q,\k,\mu) P_{11}(q)\nonumber \\
& + 2 Z_1(\mu) P_{11}(k)\left( c_\text{ct}\frac{k^2}{{ k^2_\textsc{m}}} + c_{r,1}\mu^2 \frac{k^2}{k^2_\textsc{m}} + c_{r,2}\mu^4 \frac{k^2}{k^2_\textsc{m}} \right) + \frac{1}{\bar{n}_g}\left( c_{\epsilon,0}+c_{\epsilon,1}\frac{k^2}{k_\textsc{m}^2} + c_{\epsilon,2} f\mu^2 \frac{k^2}{k_\textsc{m}^2} \right) \, , \nonumber
\end{align}
with kernels $Z_i$ defined in~\cite{Perko:2016puo,DAmico:2019fhj,Nishimichi:2020tvu}, and depending on four EFT-parameters $b_1$, $b_2$, $b_3$, $b_4$. 
Here $k_{\rm M} \equiv k_{\rm NL} = 0.7 \hinvMpc$ as defined in {the} previous section. 
In our analysis we vary the cosmological parameters $\Omega_m, h$, $\ln (10^{10}A_s)$, $n_s$ and $\omega_b$ with only a Gaussian prior on the baryon abundance $\omega_b$ motivated from Big Bang Nucleosynthesis (BBN), of width $\sigma_{\omega_b, {\rm BBN}} = 0.00036$~\cite{Mossa:2020gjc}. 
For the simulations, we will center the prior on the truth, while on BOSS data, we will use $\omega_{b,{\rm BBN}} = 0.02233$~\cite{Mossa:2020gjc}. 
We fix the neutrinos to minimal mass, $0.06 \, {\rm eV}$, as done in the Planck analysis~\cite{Planck:2018vyg}.
As for the EFT parameters, we define the linear combinations $c_2 = (b_2 + b_4) / \sqrt{2}$, $c_4 = (b_2 - b_4) / \sqrt{2}$, and we set $c_4 = 0$ since $b_2$, $b_4$ are almost completely anticorrelated.
Then we define the two combinations $c_{\epsilon, \rm mono} = c_{\epsilon,1} + f c_{\epsilon,2} / 3$, $c_{\epsilon, \rm quad} = 2 f c_{\epsilon,2} / 3$. We put a Gaussian prior of mean 0 and standard deviation 2, $\mathcal{N}(0,2)$, on $b_3$, $c_\text{ct}$, $c_{\epsilon,0}$, $c_{\epsilon, \rm mono}$, $c_{\epsilon, \rm quad}$, and a Gaussian prior of mean 0 and standard deviation 4, $\mathcal{N}(0,4)$, on the redshift-space counterterms $c_{r,1}$, $c_{r,2}$.
As explained in~\cite{DAmico:2019fhj,DAmico:2020kxu}, we analytically marginalize over these seven EFT parameters, as they appear linearly in the power spectrum and therefore quadratically in the likelihood.
Finally, the linear bias $b_1$ has a flat prior $[0, 4]$, and $c_2$ has a flat prior $[-4,4]$.

As explained in {the} previous section, we will analyze the data using two procedures: `1-loop', and `1-loop+'. 
For the `1-loop+' procedure, we will add to the theory model, Eq.~\eqref{eq:powerspectrum}, the two counterterms:
\begin{equation}
c_{r,4} \, \mu^4 \, b_1^2 \frac{k^4}{k_{\rm NL,R}^4} P_{11}, \qquad
c_{r,6} \, \mu^6 \, b_1 \frac{k^4}{k_{\rm NL,R}^4} P_{11},
\end{equation}
where $k_{\rm NL,R}^2 = k_{\rm NL}^2/8$, and we marginalize over $c_{r,4}$ and $c_{r,6}$ with a Gaussian prior centered on 0 and width 1. 

\subsection{$N$-body simulations}

Before analyzing the BOSS data, we first assess the $k$-reach of the new data combination with varying $k_{\rm max}$'s (we remind we are denoting this combination as $\PA + w_{1,2}$ and we refer to it simply as `wedges') using $N$-body simulations. 
We will stop fitting the data up to the scale where the theory-systematic error $\sigma_{\rm sys}$ stays under control. 
As done in~\cite{DAmico:2019fhj}, for each cosmological parameter, $\sigma_{\rm sys}$ is measured as the shift of the $1\sigma$-region to the truth; this can be read from the posteriors obtained fitting simulations. 
We will declare the $k_{\rm max}$ of each wedge as the maximum wavenumber of the analysis such that $\sigma_{\rm sys} < \sigma_{\rm stat}^{\rm data}/3$, where $\sigma_{\rm stat}^{\rm data}$ is the error bars of the data we want to analyze, such as BOSS.

\subsubsection{PT challenge}
We first analyze the PT challenge simulations described in~\cite{Nishimichi:2020tvu}~\footnote{More information of the PT-challenge can be found at~\url{https://www2.yukawa.kyoto-u.ac.jp/~takahiro.nishimichi/data/PTchallenge/.}}. 
The PT challenge data are the three first even multipoles of the galaxy power spectra in redshift space constructed from 10 $N$-body realizations, each of side length $3.84 h^{-1}$Gpc, sampled with $3072^3$ equal-mass particles, and populated with a BOSS-like {halo occupation distribution (HOD) model.
The measurements from each realization can be averaged into one single power spectrum and a single covariance, corresponding to a measurement in a total volume of about 100 times the volume of the BOSS DR12 sample, making the statistical error of the simulation much smaller than the one associated to any realistic galaxy {survey} at redshift $z \sim 0.6$. 

\begin{figure}[ht!]
\centering
\includegraphics[width=0.49\textwidth]{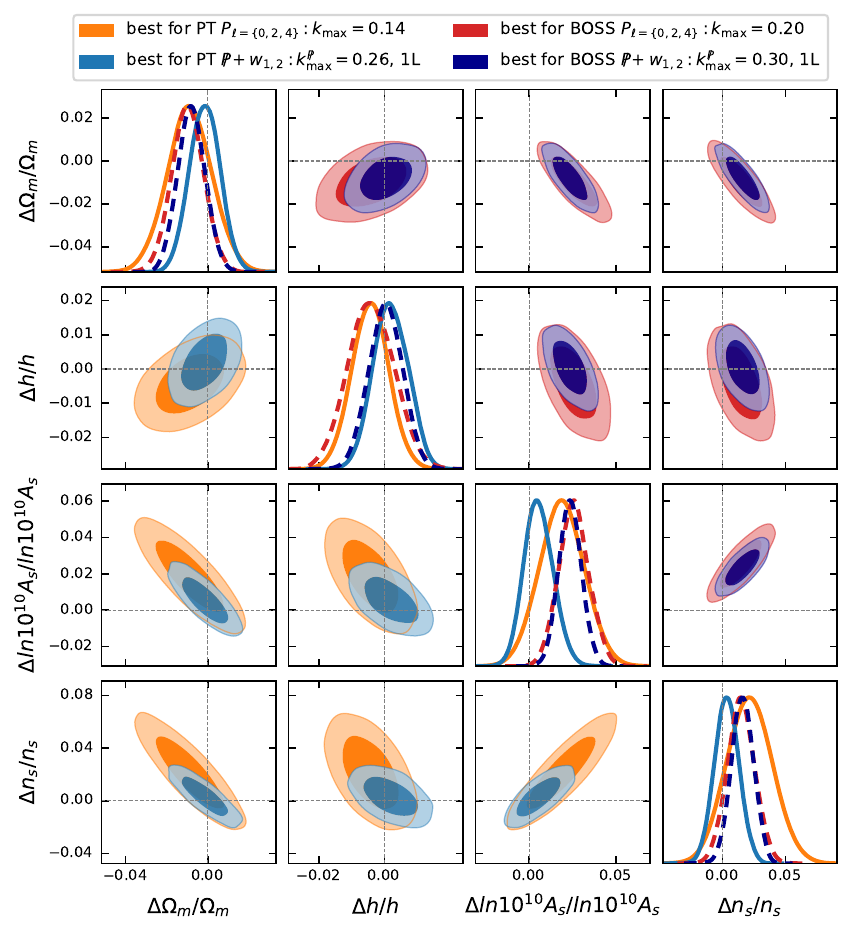}
\includegraphics[width=0.49\textwidth]{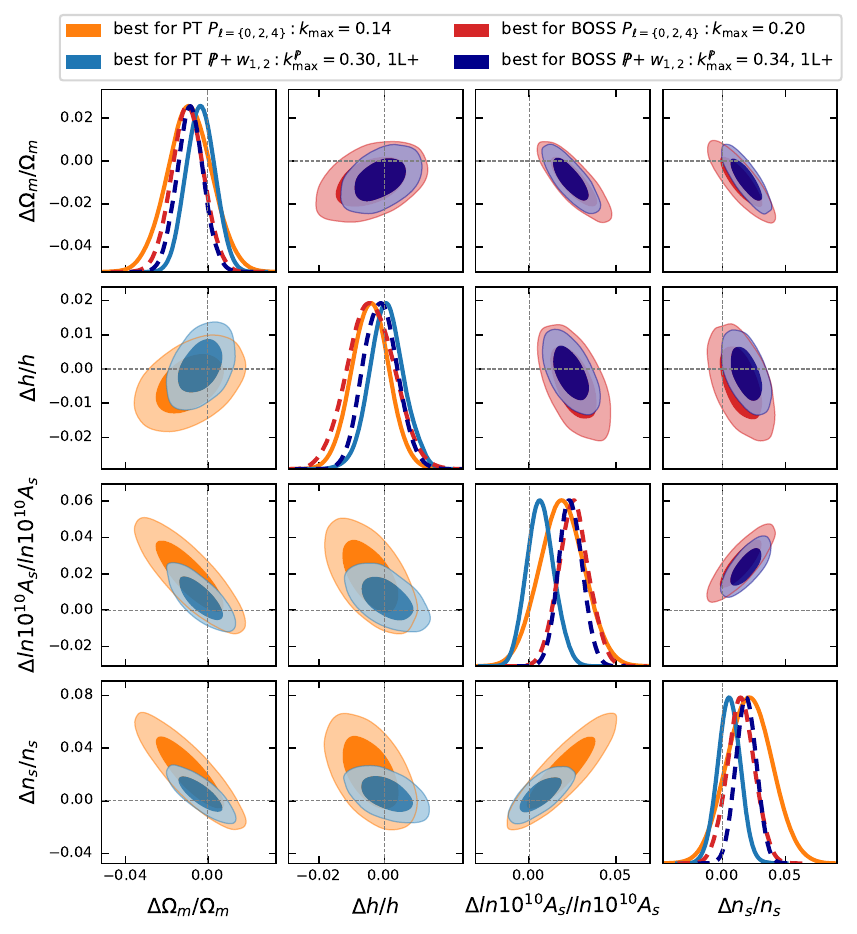}\\ \vspace{0.3em}

\scriptsize
    \begin{tabular}{|l|c|c|c|c|}
     \hline 
    $\Delta X / X$ & $P_{\ell=\{0,2,4\}}$ 0.14 			&  $\PA +  w_{1,2}$ 0.26 1L 		&  $\PA +  w_{1,2}$ 0.30 1L+ \\ \hline 
    $\Omega_{m }$ & $-0.009\pm 0.011$ 		& $-0.0014\pm 0.0075$ 			& $-0.0034\pm 0.0070$ \\ 
    $ h$ & $-0.0046\pm 0.0074$  			& $0.0019\pm 0.0053$ 			& $0.0007\pm 0.0053$ \\ 
    $\ln (10^{10}A_s)$ & $0.019\pm 0.013$  	& $0.0054^{+0.0078}_{-0.0088}$ 	& $0.0065^{+0.0074}_{-0.0082}$ \\ 
    $ n_s$ & $0.022\pm 0.018$ 				& $0.0032\pm 0.0097$			& $0.0050\pm 0.0090$ \\
    \hline 
    \end{tabular}
    \begin{tabular}{|l|c|c|c|c|}
     \hline 
    $\Delta X / X$ & $P_{\ell=\{0,2,4\}}$ 0.20 			& $\PA +  w_{1,2}$ 0.30 1L 		&  $\PA +  w_{1,2}$ 0.34 1L+ \\ \hline 
    $\Omega_{m }$ & $-0.0097\pm 0.0078$ 	& $-0.0079\pm 0.0067$ 			& $-0.0085\pm 0.0066$ \\ 
    $ h$ & $-0.0040\pm 0.0071$  			& $0.0004\pm 0.0051$ 			& $-0.0012\pm 0.0052$ \\ 
    $\ln (10^{10}A_s)$ & $0.0257\pm 0.0087$  	& $0.0235\pm 0.0066$ 			& $0.0238\pm 0.0068$ \\ 
    $ n_s$ & $0.015\pm 0.011$				& $0.0158\pm 0.0089$			& $0.0187\pm 0.0082$ \\
    \hline 
    \end{tabular}
\caption{\small Triangle plots and 68\%-confidence intervals of the cosmological parameters obtained fitting the PT challenge simulation data with a BBN prior. 
{\it Left:} Results of the fit to the multipoles $P_{\ell=\{0,2,4\}}$, up to $k_{\rm max} = 0.14$ and to the wedges $\PA + w_{1,2}$ up to $k_{\rm max}^{\PA} = 0.26$ using the `1-loop' procedure (1L), at which the cosmological parameters are recovered at best precision with a negligible theory-systematic error for the simulation volume. 
For BOSS volume, the theory-systematic error is under control up to $k_{\rm max} = 0.20$ and $k_{\rm max}^{\PA} = 0.30$, respectively.
{\it Right:} Same as on the left but using the `1-loop+' procedure (1L+) up to $k_{\rm max}^{\PA} = 0.30$ for the PT challenge, and up to $k_{\rm max}^{\PA} = 0.34$ for BOSS. 
All $k$'s are given in $\hinvMpc$.
The gray lines represent the truth of the simulation. 
{\it Bottom: } Corresponding mean and $68\%$-confidence intervals. 
} \label{fig:pt} 
\end{figure}

In Fig.~\ref{fig:pt}, we show the best results obtained fitting either the multipoles or the wedges of the PT challenge simulation, {with a BBN prior}. 
Let us first discuss the results relevant for the simulation volume. 
We will later use the PT challenge simulation to assess the $k$-reach for BOSS~\footnote{We stress that, in Fig.~\ref{fig:pt}, all chains are run assuming the PT-challenge covariance. Therefore, the results ``best for BOSS'' are not centered on the truth given such small errors. We show them as our best estimate of the theory-systematic error and $k_{\rm max}$ appropriate for a BOSS-like survey.}.

\paragraph{PT challenge results} We find that we can fit the multipoles $P_{\ell=\{0,2,4\}}$ up to $\kmax = 0.14 \hinvMpc$ with marginal theory-systematic error: $\Omega_m$ and $h$ are well recovered within $1\sigma$ and we measure relative theory-systematic errors of $43\%$ on $\ln (10^{10} A_s)$ and of $16\%$ on $n_s$~\footnote{We warn that, however, $\sigma_{\rm sys}$ is not well measured for the PT challenge simulation: we are comparing it to the statistical error measured from the same simulation, and $1\sigma$ shifts are typically {expected}, especially when considering that we are measuring $4$ parameters (and actually varying more than that). 
However, for this particular realization we find that the cosmological parameters are all well recovered within $\lesssim 1.4\sigma$ at $k_{\rm max} \leq  0.14 \hinvMpc$, which has a good $p$-value. We thus do not comment further on this. 
When compared to the statistical error obtained on other smaller-volume data, the precision of the measurements of $\sigma_{\rm sys}$ given by the $\sigma_{\rm stat}$ obtained on the PT challenge simulation is then very accurate.  }.

\begin{figure}[ht!]
\includegraphics[width=0.99\textwidth]{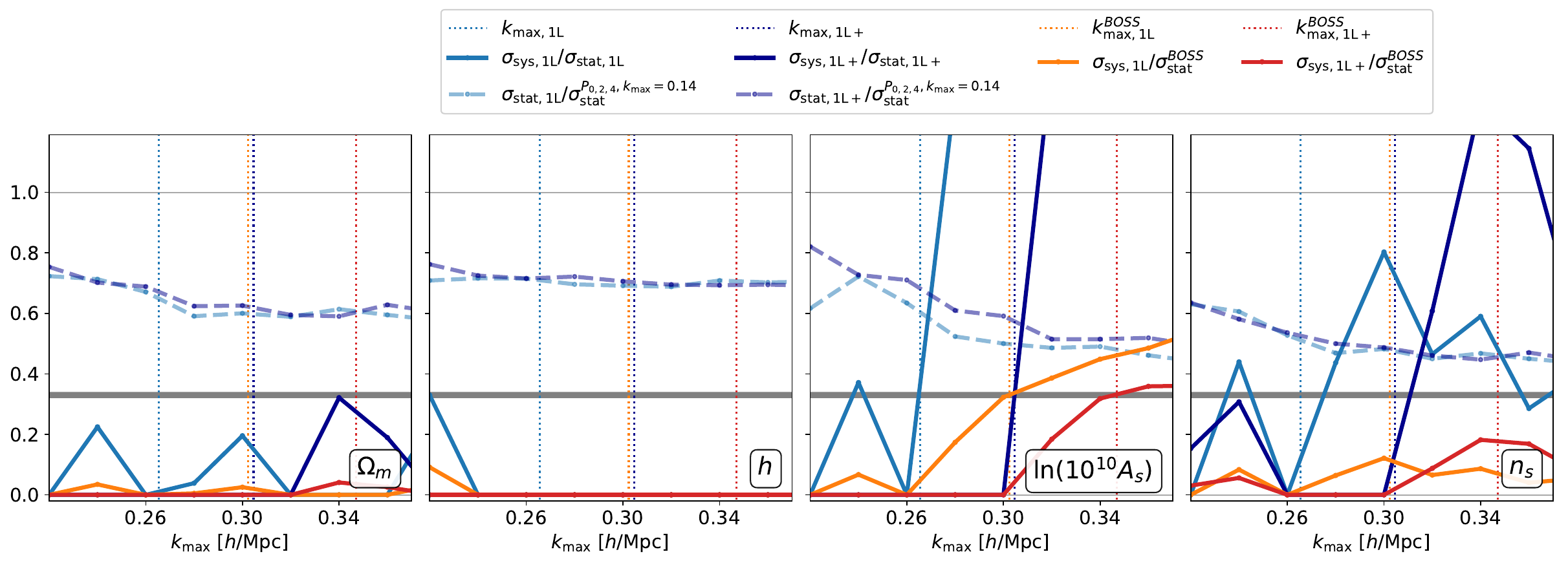}
\caption{\small In continuous line, we plot the theory-systematic error $\sigma_{\rm sys}$ relative to the statistical error $\sigma_{\rm stat}$ as a function of $k_{\rm max} \equiv k_{\rm max}^{\PA}$ measured on the PT challenge simulation fitting wedges $\PA + w_{1,2}$ using the `1-loop' procedure (1L) or the `1-loop+' procedure (1L+), with a BBN prior, for each cosmological parameter. 
The final $k_{\rm max}$ is declared at which $\sigma_{\rm sys}/\sigma_{\rm stat} \simeq 1/3$, which is represented by the grey horizontal line. 
The gain in error bars with respect to best results from the fit to multipoles $\sigma_{\rm stat}/\sigma_{\rm stat}^{P_{\ell=\{0,2,4\}}, k_{\rm max}=0.14}$ is given in dashed lines, and the best one can be read off at that $k_{\rm max}$. 
As we see, the final $k_{\rm max}$ is constrained by the $\sigma_{\rm sys}$ in $\ln(10^{10}A_s)$. 
We also show the determination of the $k_{\rm max}$ for BOSS using the same criterion $\sigma_{\rm sys} \simeq \sigma_{\rm stat}^{\rm BOSS}/3$, where $ \sigma_{\rm stat}^{\rm BOSS}$ is the error bar obtained on BOSS. 
 } \label{fig:kmax} 
\end{figure}

 For the wedges, we remind the reader that we change the $k_{\rm max}$ of $\PA$ and scale the ones of $w_{1,2}$ according to (\ref{eq:kreachbins}) or (\ref{eq:kreachbins2}). In this case, we find that we can fit the data up to much higher {wavenumbers}, substantially improving the error bars while keeping the theory-systematic error under control. 
This can be seen in Fig.~\ref{fig:kmax}, where we represent the relative size of the theory-systematic error as a function of $k_{\rm max}$ for each cosmological parameter. 
For the `1-loop' procedure, we are able to recover the cosmological parameters up to $k_{\rm max}^{\PA} = 0.26 \hinvMpc$ ($k_{\rm max}^{w_{1,2}} = 0.22, 0.11 \hinvMpc$), with no theory-systematic error. 
As it is clear from the Table in Fig.~\ref{fig:pt}, the error bars are greatly improved when going from the multipole analysis to the wedge analysis: comparing the best, theory-error-controlled, results, we find that the error bars are reduced by $33\%$ on $\Omega_m$, $29\%$ on $h$, $37\%$ on $\ln (10^{10} A_s)$, and $47\%$ on $n_s$. 
For the `1-loop+' procedure, we can fit the data up to the even higher $k_{\rm max}^{\PA} = 0.30 \hinvMpc$ ($k_{\rm max}^{w_{1,2}} = 0.21, 0.14 \hinvMpc$) with no theory-systematic error. 
Compared to the multipoles, we find that the error bars are reduced by $38\%$ on $\Omega_m$, $29\%$ on $h$, $41\%$ on $\ln (10^{10} A_s)$, and $51\%$ on $n_s$. 
This represents an improvement over the `1-loop' procedure of about $10\%$ or $15\%$ on the error bars of $\Omega_m$,  $\ln (10^{10} A_s)$, and $n_s$. 

To summarize, for the PT challenge simulation the error bars on the cosmological parameters are improved by about a factor of two thanks to our new combination of data which allows for varying $k_{\rm max}$'s. Furthermore, the `1-loop+' procedure allows us to analyze more data than the `1-loop' procedure and leads to better results of about $10\%$ or $15\%$ in error bars. 
We now turn to assess the $k$-reach for BOSS. 

\paragraph{BOSS scale cut from PT challenge} If we instead compare our measurements of the theory-systematic error with the statistical error obtained on BOSS data, we find that the theory-systematic error on the cosmological parameters stays marginally small ($\lesssim \sigma_{\rm stat}^{\rm BOSS} / 3$) up to $k_{\rm max} \sim 0.2 \hinvMpc$ for the multipoles, as found already in~\cite{DAmico:2019fhj}. 
To be precise, we find negligible $\sigma_{\rm sys}$ on $\Omega_m$, $h$ and $n_s$, while we detect a small $\sigma_{\rm sys} = 0.30 \sigma_{\rm stat}^{\rm BOSS}$ on $\ln (10^{10} A_s)$ at $k_{\rm max} = 0.2 \hinvMpc$. 
For the wedges, as it can be seen from Fig.~\ref{fig:kmax}, we find that we can fit the data up to $k_{\rm max}^{\PA} = 0.30 \hinvMpc$ ($k_{\rm max}^{w_{1,2}} = 0.25, 0.12 \hinvMpc$) for the `1-loop' procedure, and up to $k_{\rm max}^{\PA} = 0.34 \hinvMpc$ ($k_{\rm max}^{w_{1,2}} = 0.24, 0.16 \hinvMpc$) for the `1-loop+' procedure. 
The theory-systematic error again essentially accumulates in $\ln (10^{10} A_s)$, where it remains safely small ($<\sigma_{\rm stat}^{\rm BOSS} / 3$), at the $k_{\rm max}$'s we quote. Here the theory-systematic error is measured with a precision which is given by the errors bars obtained on the PT challenge simulation, that can be read off from the Table in Fig.~\ref{fig:pt}: in terms of BOSS error bars, the theory-systematic error is detected with a precision of $\sim 15\%$ on $\Omega_m$, $\ln(10^{10}A_s)$, and $n_s$, and $\sim 25\%$ on $h$. 
These represent small corrections to our BOSS results, as these uncertainties becomes negligible once added in quadrature to the error budget. 

\paragraph{} In order to obtain a robust estimate of the $k$-reach, it is important to marginalize over the the `micro-physics' of the simulations, such as e.g. the choice of the HOD model, by analyzing many of them. 
We thus now move to another set of simulations. 

\subsubsection{Lettered challenge}
The `lettered' challenge boxes are a set of high-fidelity simulations described in~\cite{DAmico:2019fhj} that we already used to assess the scale cut of the EFTofLSS when analyzing multipoles. 
The lettered simulations consist in two independent realizations of side length $2.5 \, {\rm Gpc}/h$. 
One is populated by 4 different HOD models, labelled A, B, F, G, and the other one, labelled D, is populated by a different HOD model.
Details on the HOD models and other specifics of those simulations can be found in~\cite{Alam:2016hwk}.
For BOSS, we found that we can fit the multipoles up to $k_{\rm max} = 0.23 \hinvMpc$ when analyzing the data with a BBN prior. 
We now do the same study for the new data combination, the wedges, with varying $k_{\rm max}$'s. 
{We proceed in the following way}. 
We fit all boxes separately and average the posteriors of the cosmological parameters obtained on boxes A, B, F, and G, as they are from the same seed. 
We call this combination ABFG. 
As D is an independent realization, we can combine its results with the ones of ABFG, allowing us to measure the theory-systematic error with a better precision of about $\sqrt{2}$. 
To do so, we combine the individual 1D posteriors of the shifts from the truth of ABFG with the ones of D, as the product of two Gaussians. 
By comparing the resulting shifts of the $1\sigma$ region from the truth, to the error bars obtained on BOSS, we get another precise measurement of the theory-systematic error. 

\begin{figure}[ht!]
\centering
\includegraphics[width=0.49\textwidth]{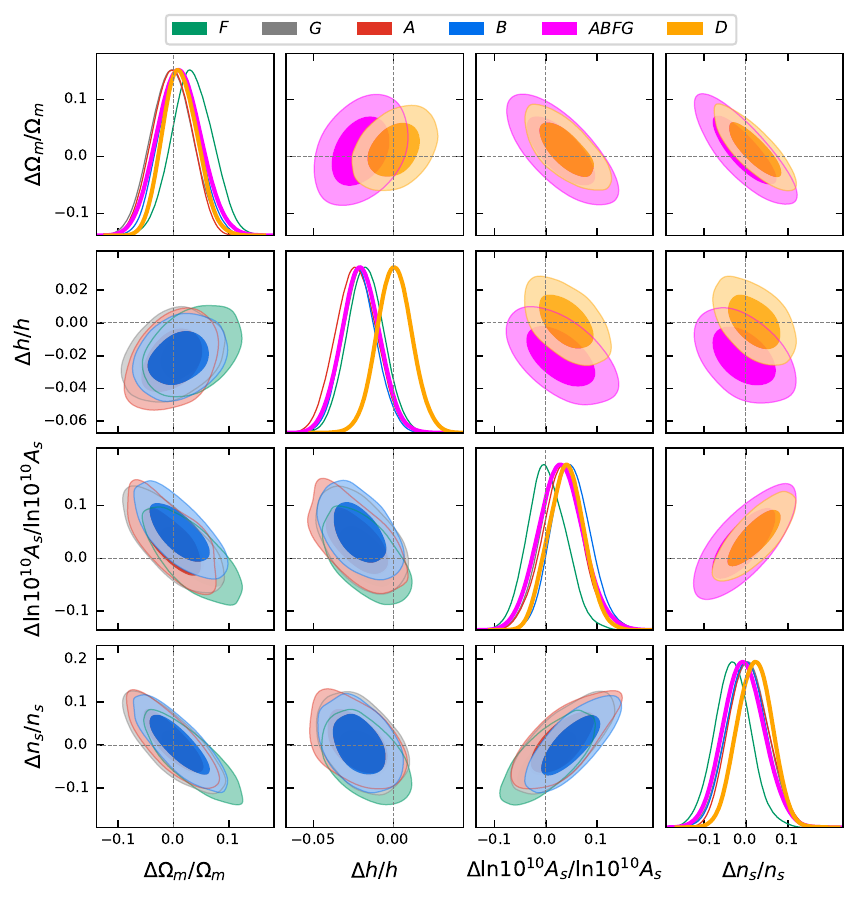}
\includegraphics[width=0.49\textwidth]{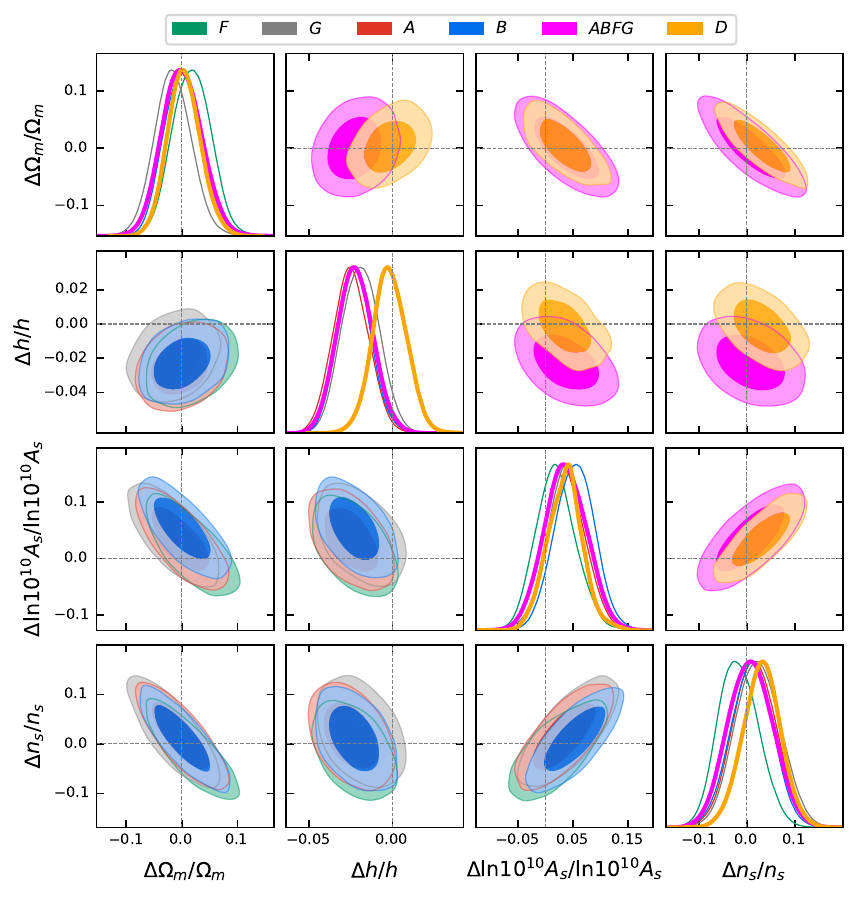}\\ \vspace{0.3em}

\scriptsize
    \begin{tabular}{|l|c|c|c|c|}
     \hline 
    $\Delta X / X$ & $P_{\ell=\{0,2,4\}}$ 0.23 			& $\PA +  w_{1,2}$ 0.26 1L 		&  $\PA +  w_{1,2}$ 0.32 1L+ \\ \hline 
    $\Omega_{m }$ & $0.003\pm 0.022$ 	& $0.011\pm 0.024$ 			& $-0.003\pm 0.023$ \\ 
    $ h$ & $-0.008\pm 0.007$  			& $-0.009\pm 0.008$ 			& $-0.011\pm 0.007$ \\ 
    $\ln (10^{10}A_s)$ & $0.047\pm 0.024$  	& $0.036\pm 0.027$ 			& $0.036\pm 0.025$ \\ 
    $ n_s$ & $0.029\pm 0.029$				& $0.010\pm 0.031$			& $0.022\pm 0.030$ \\
    \hline 
    \end{tabular}

\caption{\small Triangle plots and $68\%$-confidence intervals of the cosmological parameters obtained fitting the lettered challenge simulation data with a BBN prior. 
{\it Left:} Results of the fit to the wedges $\PA + w_{1,2}$ up to $k_{\rm max}^{\PA} = 0.26$ using the `1-loop' procedure (1L), at which the cosmological parameters are recovered at best precision with a negligible theory-systematic error compared to BOSS error bars. 
{\it Right:} Same as on the left but using the `1-loop+' procedure (1L+) up to $k_{\rm max}^{\PA} = 0.32$. 
All $k$'s are given in $\hinvMpc$. 
The gray lines represent the truth of the simulation. 
{\it Bottom: } Corresponding mean and $68\%$-confidence intervals. Here we quote the most precise combination, ABFG+D, as described in the main text. 
} \label{fig:letter} 
\end{figure}

\paragraph{BOSS scale cut from lettered challenge} In Fig.~\ref{fig:letter}, we show the best results fitting the lettered challenge simulation data with a BBN prior. 
Results fitting two multipoles can be found in~\cite{Colas:2019ret} (see also~\cite{DAmico:2020kxu,DAmico:2020tty} for the multipole analysis of these simulations in other cosmologies). 
We find that the theory-systematic error is marginally small up to $k_{\rm max}^{\PA} = 0.26 \hinvMpc$ ($k_{\rm max}^{w_{1,2}} = 0.22, 0.10 \hinvMpc$) for the `1-loop' procedure, and up to $k_{\rm max}^{\PA} = 0.32 \hinvMpc$ ($k_{\rm max}^{w_{1,2}} = 0.23, 0.15 \hinvMpc$) for the `1-loop+' procedure: relatively to BOSS error bars, we measure small theory-systematic errors from the combination ABFG+D of about, respectively, $5\%$ and $20\%$ in $h$, $15\%$ and $19\%$ in $\ln (10^{10} A_s)$, and none in $\Omega_m$ or $n_s$. 
In terms of BOSS error bars, here the theory-systematic error is detected with an uncertainty of $\sim 40\%$ for all cosmological parameters. 
Again, this represents a negligible correction to our BOSS results, as these uncertainties count only for $\sim 10\%$ once added in quadrature to the error budget. 
At higher $k_{\rm max}$'s, the theory-systematic error in $\ln (10^{10} A_s)$ becomes statistically significant. 
The $k_{\rm max}$'s that we get from the lettered challenge simulations are slightly {smaller} by a few $k$ bins ($\Delta k \sim 0.01$) than the answer that we get from the PT challenge simulation. 
To stay on the safe side, we will choose the most conservative choice to analyze the BOSS data. 

\paragraph{} To summarize, from simulations we learn that we can {confidently} fit the BOSS data in wedges up to $k_{\rm max}^{\PA} = 0.26 \hinvMpc$ ($k_{\rm max}^{w_{1,2}} = 0.22, 0.10 \hinvMpc$) for the `1-loop' procedure, and up to $k_{\rm max}^{\PA} = 0.32 \hinvMpc$ ($k_{\rm max}^{w_{1,2}} = 0.23, 0.15 \hinvMpc$) for the `1-loop+' procedure, with a controlled theory-systematic error.

\subsection{BOSS}
As {we did} for the simulations, we use multipole measurements that we eventually rotate to our new data combination as described in Subsection~\ref{sec:lkl}. 
The power spectrum multipole measurements are the ones obtained and described in~\cite{Zhang:2021yna}~\footnote{Contrary to the analysis of~\cite{Zhang:2021yna} that follows~\cite{DAmico:2019fhj,Colas:2019ret}, we remind that in this work we analyze three multipoles instead of two (or wedges constructed from three multipoles), fix the neutrinos to minimal mass, and do not set the stochastic term $\sim k^2$ in the monopole to zero. }. 
We fit $4$ skycuts, CMASS NGC, CMASS SGC, LOWZ NGC, and LOWZ SGC, {assigning to} each an independent set of EFT parameters. 
These skycuts are constructed from the BOSS DR12 catalogs~\cite{Reid:2015gra} given the redshift range $0.2 < z< 0.43$  for LOWZ and $0.43 < z < 0.7$ for CMASS.
The covariances are obtained from 2048 ``Patchy mocks'' described in~\cite{Kitaura:2015uqa}.
All observational effects such as window functions, etc. are accounted as described in~\cite{DAmico:2019fhj} and using \code{PyBird}~\cite{DAmico:2020kxu}. 

\begin{figure}[ht!]
\centering
\includegraphics[width=0.87\textwidth]{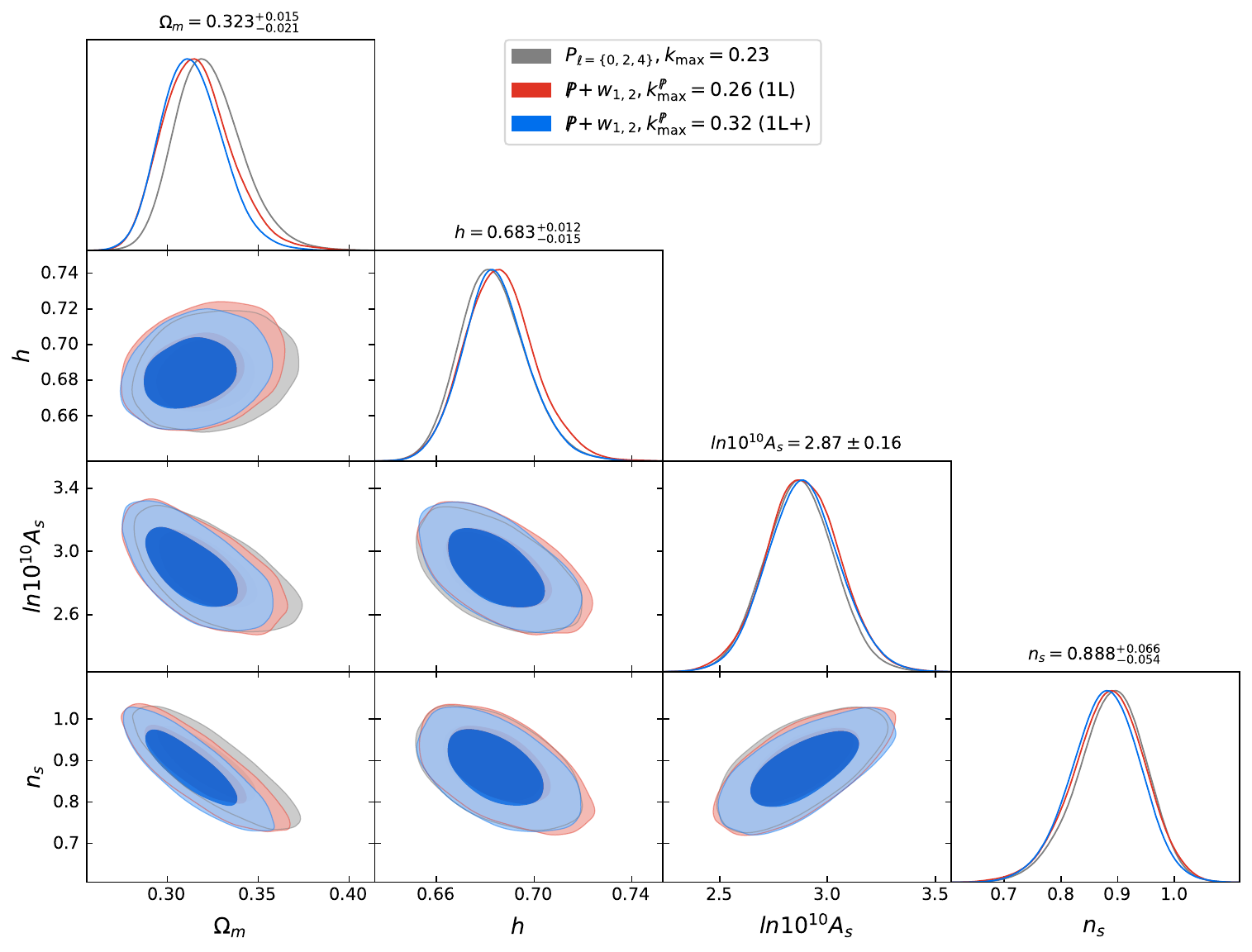}  \\ \vspace{0.3em}
\footnotesize
    \begin{tabular}{|l|c|c|c|c|}
     \hline 
    ${\rm Params}$ 		& $P_{\ell=\{0,2,4\}}$ 0.23 				& $\PA +  w_{1,2}$ 0.26 1L 		&  $\PA +  w_{1,2}$ 0.32 1L+ \\ \hline 
    $\Omega_{m }$ 	& $0.323^{+0.015}_{-0.021}$ 		& $0.317^{+0.015}_{-0.021}$		& $0.314^{+0.015}_{-0.019} $ \\ 
    $ h$ 			& $0.683^{+0.012}_{-0.015}$ 		& $0.686^{+0.013}_{-0.015}$		& $0.684^{+0.012}_{-0.014}   $ \\ 
    $\ln (10^{10}A_s)$& $2.87\pm 0.16$  				& $2.89\pm 0.17 $ 				& $2.89\pm 0.17              $\\ 
    $ n_s$ 			& $0.888^{+0.066}_{-0.054}  $		& $0.884^{+0.067}_{-0.058}  $		& $0.878^{+0.063}_{-0.057}   $ \\
    \hline 
    \end{tabular}
    
\caption{\small Triangle plots and $68\%$-confidence intervals of the cosmological parameters obtained fitting the multipoles or the wedges of BOSS data with a BBN prior. 
All $k$'s are given in $\hinvMpc$. 
} \label{fig:boss} 
\end{figure}

In Fig.~\ref{fig:boss}, we show the results obtained fitting the wedges of BOSS data with a BBN prior. 
The improvement in the error bars is marginal: while we do not gain from the `1-loop' procedure, from the `1-loop+' procedure we gain about $10\%$ in $\Omega_m$, while the error bars on $h$, $\ln(10^{10}A_s)$, and $n_s$ are similar at $\lesssim 3\%$. 

Let us add a comment on the values of the $c_{r,4}$, $c_{r,6}$ parameters in the `1-loop+' procedure. We checked analyzing the single CMASS NGC sky that their values is consistent with zero at 1-$\sigma$, and in particular $c_{r,6}$ is completely prior-dominated.
This shows that we are safely in the perturbative regime. As we increase $k_{\rm max}^{\slashed{P}}$, they start to deviate from zero as soon as the $k_{\rm max}^{w_1}$ (obtained by eq.~\eqref{eq:kreachbins2}) becomes comparable to $k_{\rm NL,R}$. In fact, at this point we expect that perturbation theory starts to converge very slowly.

Thus, the gain in BOSS is much less than the gain in the PT challenge as we go from multipoles to wedges. 
We have checked on the PT challenge simulation by rescaling the covariance to the volume of BOSS data or to an intermediate volume, or by using Fisher matrix, that indeed, the gain in error bars depends on the data volume, {\it i.e.} on the cosmic variance. 
We now turn to see what improvements we can expect for the next-stage surveys such as DESI. 

\begin{figure}[ht!]
\centering
\includegraphics[width=0.49\textwidth]{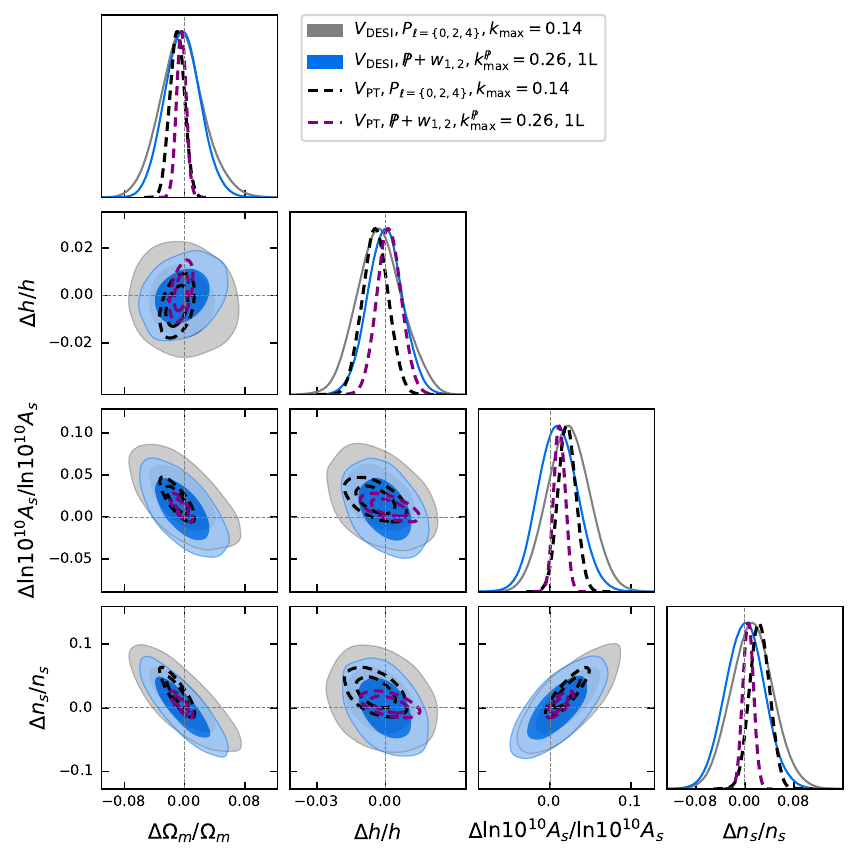}
\includegraphics[width=0.49\textwidth]{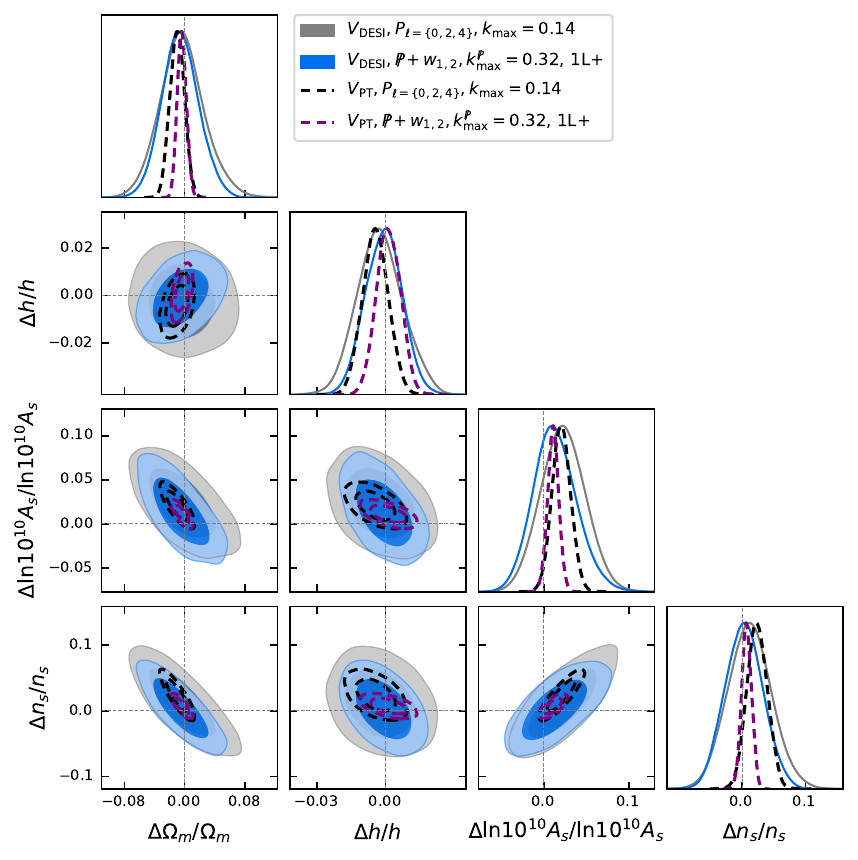}

\scriptsize
    \begin{tabular}{|l|c|c|c|c|}
     \hline 
    $\Delta X / X$ & $P_{\ell=\{0,2,4\}}$ 0.14 				&  $\PA +  w_{1,2}$ 0.26 1L 		&  $\PA +  w_{1,2}$ 0.32 1L+ \\ \hline 
    $\Omega_{m }$ & $-0.003^{+0.027}_{-0.030} $ 	& $-0.003\pm 0.024 $ 			& $-0.006\pm 0.024           $ \\ 
    $ h$ & $-0.0023^{+0.0093}_{-0.011}$  			& $-0.0002\pm 0.0076  $ 			& $-0.0008\pm 0.0080         $\\ 
    $\ln (10^{10}A_s)$ & $0.023\pm 0.026$  		& $0.0096\pm 0.024 			$ 	& $0.013^{+0.023}_{-0.027}   $ \\ 
    $ n_s$ & $0.012\pm 0.034 $					& $0.000\pm 0.032  $			& $0.003\pm 0.030            $\\
    \hline 
    \end{tabular}
    
\caption{\small Triangle plots and $68\%$-confidence intervals of the cosmological parameters obtained fitting the PT challenge simulation data with a BBN prior and DESI covariance. 
{\it Left:} Results of the fit to the multipoles $P_{\ell=\{0,2,4\}}$, up to $k_{\rm max} = 0.14$, and to the wedges $\PA + w_{1,2}$ up to $k_{\rm max}^{\PA} = 0.26$ using the `1-loop' procedure (1L).
At these scale cuts, the cosmological parameters are recovered at best precision with a marginally-small theory-systematic error $\sigma_{\rm sys}/\sigma_{\rm stat}^{\rm DESI} < 1/3$. 
$\sigma_{\rm sys}$ is measured using the covariance corresponding to the PT challenge volume $V_{\rm PT}$, while the error bars $\sigma_{\rm stat}^{\rm DESI}$ obtained using the covariance rescaled to DESI volume $V_{\rm DESI}$. 
{\it Right:} Same as on the left but using the `1-loop+' procedure (1L+) up to $k_{\rm max}^{\PA} = 0.32$. 
All $k$'s are given in $\hinvMpc$. 
The gray lines represent the truth of the simulation. 
{\it Bottom: } Corresponding mean and $68\%$-confidence intervals with the DESI covariance rescaled to DESI volume.  
} \label{fig:desi} 
\end{figure}

\subsection{DESI-like survey}
Here we will argue that there is a substantial gain to analyze the data with our wedges combination instead of multipoles for a survey like DESI. 
To show this, we use the PT challenge simulation as follows. 
First, we re-compute the covariance assuming Gaussianity as described in~\cite{Nishimichi:2020tvu} but with galaxy number density $\bar{n}_g^{\rm DESI} = 2 \times 10^{-3} \, h^3/\textrm{Gpc}^3$, which is the expected {one for} DESI. 
We then analyze the PT challenge multipoles and wedges with this new, `DESI' covariance. 
We also rescale the prior of the constant stochastic bias in Eq.~\eqref{eq:powerspectrum} with $\bar{n}_g^{\rm DESI}$, but not the ones of the stochastic terms going as $\sim k^2$, as only the constant shot noise is subtracted from the simulation data. 
Using the total simulation volume, we can get a good measurement of the theory-systematic error, while if we rescale the DESI covariance to the volume $25.5 \,(\textrm{Gpc}/h)^3$, we can get a good estimate of the error bars we can expect to obtain with DESI. 
Notice that the PT challenge were calibrated for BOSS selection functions and specifics. 
We thus warn that our present study for DESI is approximate. 
However, we find that the conclusions we reach here are confirmed by Fisher matrix study, in which we can put the right priors on all stochastic terms. 
In particular, the gain in error bars stays close to the numbers we quote if we vary the redshift from $z_{\rm BOSS}=0.61$ to $z_{\rm DESI} \sim 1$. 
Therefore, we are confident in the error bars we quote, and, we have checked that our conclusions are mostly unaffected as we vary reasonably the scale cut around the one we estimate here. 

\paragraph{} In Fig.~\ref{fig:desi}, we show the best results obtained fitting the multipoles and wedges with a BBN prior of the PT challenge simulation with the DESI covariance. 
For a theory-systematic error no larger than $<\sigma_{\rm stat}^{\rm DESI} /3$, the multipoles can be fit up to $k_{\rm max} = 0.14 \hinvMpc$, while the wedges can be fit up to $k_{\rm max}^{\PA} = 0.26 \hinvMpc$ ($k_{\rm max}^{w_{1,2}}= 0.22, 0.11 \hinvMpc$) for the `1-loop' procedure, and up to $k_{\rm max}^{\PA} = 0.32 \hinvMpc$ ($k_{\rm max}^{w_{1,2}}= 0.23, 0.15\hinvMpc$) for the `1-loop+' procedure. 
This can be seen from Fig.~\ref{fig:desi_kmax}. 
The error bars obtained by fitting the wedges with both procedures are similar ($\lesssim 5\%$ difference), and shrink by about $20\%$ on $\Omega_m$ and $h$, $5\%$ on $\ln(10^{10}A_s)$, and $10\%$ on $n_s$, compared to the multipole results. 

\begin{figure}[]
\includegraphics[width=0.99\textwidth]{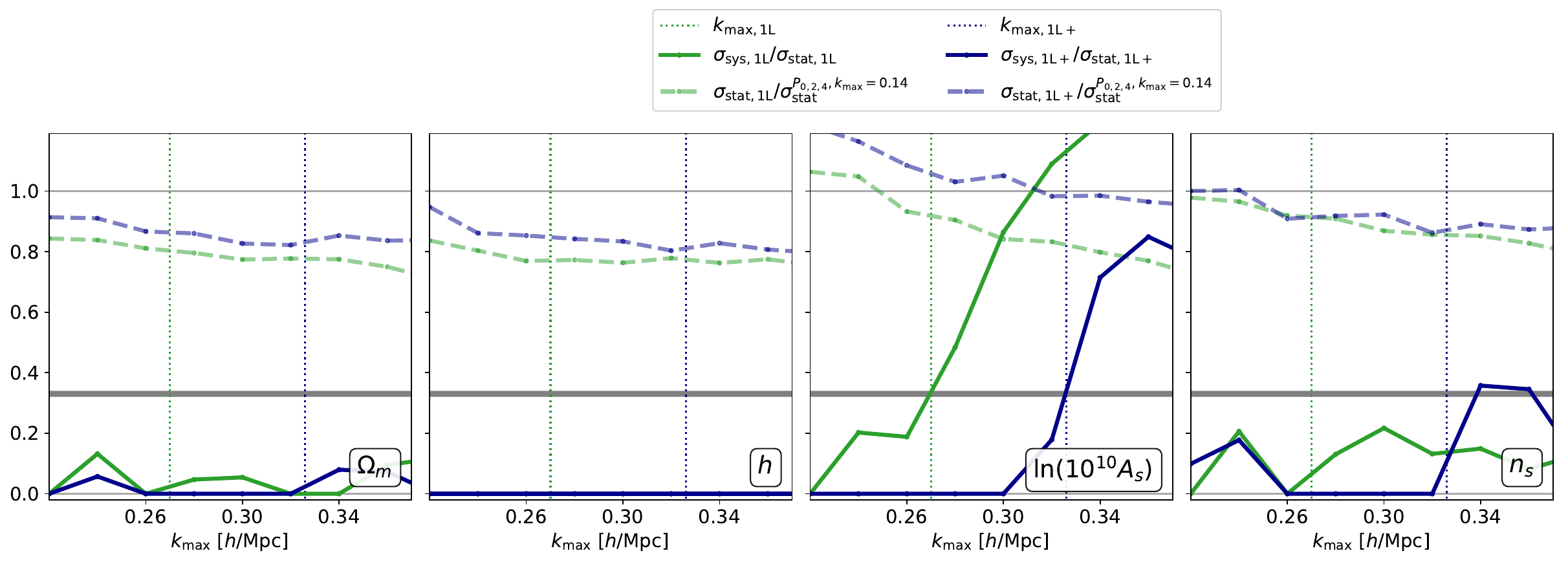}
\caption{\small In continuous line, we plot the theory-systematic error $\sigma_{\rm sys}$ relative to the statistical error $\sigma_{\rm stat}$ as a function of $k_{\rm max} \equiv k_{\rm max}^{\PA}$ measured on the PT challenge simulation with the DESI covariance, fitting wedges $\PA + w_{1,2}$ using the `1-loop' procedure (1L) or the `1-loop+' procedure (1L+), with a BBN prior, for each cosmological parameter. 
$\sigma_{\rm sys}$ is measured using the covariance corresponding to the full PT challenge volume, while $\sigma_{\rm stat}$ is measured using the covariance rescaled at DESI volume. 
The final $k_{\rm max}$ is declared at which $\sigma_{\rm sys}/\sigma_{\rm stat} \simeq 1/3$, which is represented by the grey horizontal line. 
The gain in error bars with respect to best results from the fit to multipoles $\sigma_{\rm stat}/\sigma_{\rm stat}^{P_{\ell=\{0,2,4\}}, k_{\rm max}=0.14}$ is given in dashed lines, and the best one can be read off at that $k_{\rm max}$. 
 } \label{fig:desi_kmax} 
\end{figure}

\section*{Acknowledgments}
The data analysis was performed in part on the Sherlock cluster at the Stanford University, for which we thank the support team, in part on the computer clusters LINDA \& JUDY in the particle cosmology group at the University of Science and Technology of China, and in part on the HPC (High Performance Computing) facility of the University of Parma, Italy.
This work was supported in part by MEXT/JSPS KAKENHI Grant Number JP19H00677, JP20H05861 and JP21H01081.
We also acknowledge financial support from Japan Science and Technology Agency (JST) AIP Acceleration Research Grant Number JP20317829.
The simulation data analysis was performed partly on Cray XC50 at Center for Computational Astrophysics, National Astronomical Observatory of Japan.

\begin{appendix}
\end{appendix}

\bibliographystyle{JHEP}
\bibliography{references}

\end{document}